\documentclass[english,nofootinbib,notitlepage]{revtex4}

\usepackage[T1]{fontenc}
\usepackage[latin9]{inputenc}
\usepackage{color}
\usepackage{graphicx}
\usepackage{amssymb,amsmath}
\usepackage{babel}

\def\be{\begin{equation}}
\def\ee{\end{equation}}

\makeatletter

\@ifundefined{definecolor}
 {\usepackage{color}}{}

\begin{document}

\title{Higgs Mass and Gravity Waves in Standard Model False Vacuum Inflation}

\author{Alessio Notari$^{1,2}$}
\email{notari@fe.infn.it}

\affiliation{$^{1}$  Dip.~di Fisica, Universit\`a di Ferrara and INFN Sez.~di Ferrara, Via Saragat 1, I-44100 Ferrara, Italy}
\affiliation{$^{2}$ Departament de F\'isica Fondamental i Institut de Ci\`encies del Cosmos, Universitat de Barcelona, Mart\'i i Franqu\`es 1, 08028 Barcelona, Spain}

\begin{abstract}
In previous publications we have proposed that Inflation can be realized in a second minimum of the Standard Model Higgs potential at  energy scales of about $10^{16}$ GeV, if the minimum is not too deep and if a mechanism which allows a transition to the radiation dominated era can be found. This is provided, {\it e.g.}, by scalar-tensor gravity models or hybrid models. Using such ideas we had predicted the Higgs boson mass  to be of about $126\pm 3$ GeV, which has been confirmed by the LHC, and that a possibly measurable amount of gravity waves should be produced.
Using more refined recent theoretical calculations of the RGE we show that such scenario has the right scale of Inflation only for small Higgs mass, lower than about 124 GeV, otherwise gravity waves are overproduced. The precise value is subject to some theoretical error and to experimental errors on the determination of the strong coupling constant.
Such an upper bound corresponds also to the recent claimed measurement by BICEP2 of the scale of inflation through primordial tensor modes. Finally we show that introducing a moderately large non-minimal coupling for the Higgs field the bound can shift to larger values and be reconciled with the LHC measurements of the Higgs mass.
\end{abstract}


 \maketitle



In~\cite{Masina:2011aa,Masina:2011un, Masina:2012yd} we have proposed that Inflation in the Early Universe can be realized for a narrow band of values of the top quark and Higgs boson masses, for which
the Standard Model (SM) Higgs potential develops a second local minimum~\cite{CERN-TH-2683, hep-ph/0104016, strumia2, Bezrukov:2012sa, strumia3} at energy scales of about $10^{16}$ GeV.
Such a  scenario is viable if a successful transition to a radiation-dominated era can be obtained, which we have shown to be possible in ref.~\cite{Masina:2011aa}  using an explicit model in the framework of a scalar-tensor theory of gravity developed in~\cite{hep-ph/0511207,astro-ph/0511396} and in~\cite{Masina:2012yd} using a hybrid inflationary model with an additional weakly coupled scalar, in standard gravity. Such a generic scenario could be be realized only for a Higgs mass $m_H$  in the range $m_H=(126.0 \pm 3.5)$ GeV, the error being mainly due to the theoretical uncertainty of the 2-loops Renormalization Group Equations (RGE)  used in that calculation. Such a prediction for the Higgs mass range has turned out to be surprisingly compatible with the measurement of a Higgs mass of $125.9\pm 0.4$ GeV by the LHC~\cite{HCP11,TEV}.

As it is well known, Inflation can generate quantum mechanically tensor (gravity wave) modes, usually parameterized
through $r$, the ratio of tensor-to-scalar perturbation spectra at large scales. In~\cite{Masina:2011aa, Masina:2011un} we had predicted a  relatively large amplitude for $r$ in our scenario, which is set by the overall height of the second minimum, which can be computed up to the precision of the RGE calculations and of the experimental input for the Standard Model parameters: $m_H$, the top mass $m_t$ and the strong coupling constant $\alpha_s$. However the situation has recently changed, since more refined theoretical calculations of the Standard Model potential have been performed, therefore making more precise the correspondence between the input parameters and the scale of the second minimum.
Moreover the tensor modes have now been claimed to be measured by the BICEP2 collaboration~\cite{BICEP2} with a large amplitude, which corresponds to a scale of about $2\times 10^{16}$ GeV.

In this paper we update our results,  using the results of~\cite{strumia2,strumia3} for the Higgs effective potential, and show that if the theoretical errors are under control the scale of Inflation is generically larger than $2\times 10^{16}$ GeV and can be reconciled with experiments only if $m_H\lesssim 124$ GeV (plus a theoretical error, at present estimated to be of order $0.3$ GeV).
Of course such conclusion can be changed if extra ingredients are added to the Standard Model. We show the results that can be obtained by introducing a non-minimal coupling of the Higgs to gravity. Of course less minimal modifications, such as a direct coupling of the Higgs to an extra scalar, can also easily shift the scale of Inflation. 
Finally, provided the scale of Inflation due to the Higgs potential is low enough, we also reanalyze the specific model  based on scalar tensor theories~\cite{Masina:2011aa}, which realizes the exit to a radiation dominated era, showing that  it could reproduce  observations on $r$ and on the scalar spectral index $n_s$.


As in~\cite{Masina:2011aa,Masina:2011un,Masina:2012yd} we consider the  potential for the Higgs field $\chi$ in the SM of particle physics.
For very large values of the Higgs field such potential can be written as 
\begin{equation}
V(\chi) \simeq \lambda_{\rm eff}(\chi)/4 \, \chi^4\,\,.
\end{equation}
where $\lambda_{\rm eff}$ is some effective quartic coupling~\cite{strumia2} which is very close, but not identical, to the quartic coupling $\lambda$.  Due to the running of couplings $\lambda_{\rm eff}$ can become very small at high energies, while keeping always positive values (stability regime). In such conditions the potential can have an additional minimum at very high field values. While this is at present disfavored by the current calculations and measurements in the SM it still not ruled out. In particular this can still happen if the top quark mass has a  low value compared to the present best fit~\cite{PDG,ATLAS:2014wva}.

If the Higgs field starts in the false minimum at  $\chi_0$ and dominates the energy density of the Universe, the Friedmann equation leads to a stage of inflationary expansion 
\be
H^2  \simeq \frac{V(\chi_0)}{3 M^2} \equiv H_I^2 \,\,\,,\,\,\,  \,\,a(t)\propto e^{H_I t} 
\label{eq-M}
\ee
where  $a(t)$ is the scale factor, $H \equiv \dot a /a$ is the Hubble rate and $M$ is the reduced Planck mass ($M\equiv 2.435 \times 10^{18} GeV$).

A nontrivial model-dependent ingredient is a mechanism to achieve a graceful exit from Inflation, that is a transition to a radiation-dominated era, in a nearly flat Universe at a sufficiently high-temperature. 
In~\cite{Masina:2011aa} we have proposed that the Higgs field can tunnel to the other side of the potential barrier by nucleating bubbles~\cite{Coleman} 
that can successfully collide and percolate, in the presence of  a scalar tensor theory of gravity. Alternatively in~\cite{Masina:2012yd} we had proposed that a smooth transition may happen also in standard gravity, in the presence of an extra scalar field very weakly coupled to the Higgs field. The main point of both mechanism is that they do {\it not} affect appreciably the SM runnings of couplings, since in the first model there is only a gravitational coupling to the Standard Model, and in the second case the dimensionless coupling constant is extremely small, of order $10^{-11}$. For this reason precise connections with low energy parameters still hold even in the presence of such new fields. Of course it is possible that such a new scalar can  also interact with  the Higgs field with a large direct coupling and in this case this could also affect the RGE equations and we briefly discuss later such possibilities.

Subsequently the Higgs field could roll down the potential, reheat the Universe and finally relax in the present true vacuum with $v=246$ GeV.
 In the scalar-tensor theory of gravity model an extra  scalar field $\phi$ is introduced, the Brans-Dicke scalar or dilaton, which has an interaction term of the form
$f(\phi) R$,  where $R$ is the Ricci scalar and $f(\phi)>0$ thus sets the value of the Planck mass. The presence of such field makes the Planck mass time-dependent, and therefore also $H$ during Inflation: this slows down the expansion sufficiently enough if $f(\phi)$ grows faster than $\phi^2$ for large $\phi$ values, since a stage of quasi exponential expansion is followed by a stage of power-law (even decelerated) expansion ~\cite{hep-ph/0511207,astro-ph/0511396}. 
During the exponential phase  the quantum fluctuations in $\phi$ lead to a nearly scale-invariant spectrum of perturbations. 
During the subsequent decelerated phase, $H$ decreases rapidly, and therefore  the expansion is sufficiently slowed down, so that many bubbles can be nucleated in a Hubble patch and subsequently collide and reheat the Universe, leading to a spatially flat radiation dominated Universe.

As discussed in~\cite{Masina:2011aa}, after tunneling we require the field $\phi$ to relax to zero if a suitable potential $U(\phi)$ is present, which allows us to identify the present reduced Planck mass with the quantity $M$
and, at the same time, to satisfy constraints from fifth-force experiments and time-dependence of the Newton constant $G_N=1/(8 \pi M^2)$  ~\cite{Will:2005va}.

An alternative to scalar-tensor theories is given by models~\cite{Masina:2012yd}
with a direct tiny coupling of the Higgs field to an additional scalar field,
which induces a time-dependence directly into $\Gamma$ by flattening the barrier in the potential. 

It is crucial to notice that  a graceful exit can be  realized in the above models only if  there is a very shallow false minimum, otherwise the tunneling rate would be negligibly small in the scalar-tensor model, since the probability is exponentially sensitive to the barrier~\cite{Coleman}. Such shallowness is required also in the hybrid model since  a change in the barrier by a large amount would introduce a departure from an almost flat scalar spectral index.
So, the shape of the potential is very close to the case in which there is just an inflection point and thus we have a generic prediction for the scale of Inflation and for $r$, while the specific  model only affects the prediction for the spectral index of cosmological density perturbations $n_S$.

\begin{figure}[t!]
 \centering 
 \includegraphics[width=8.5cm]{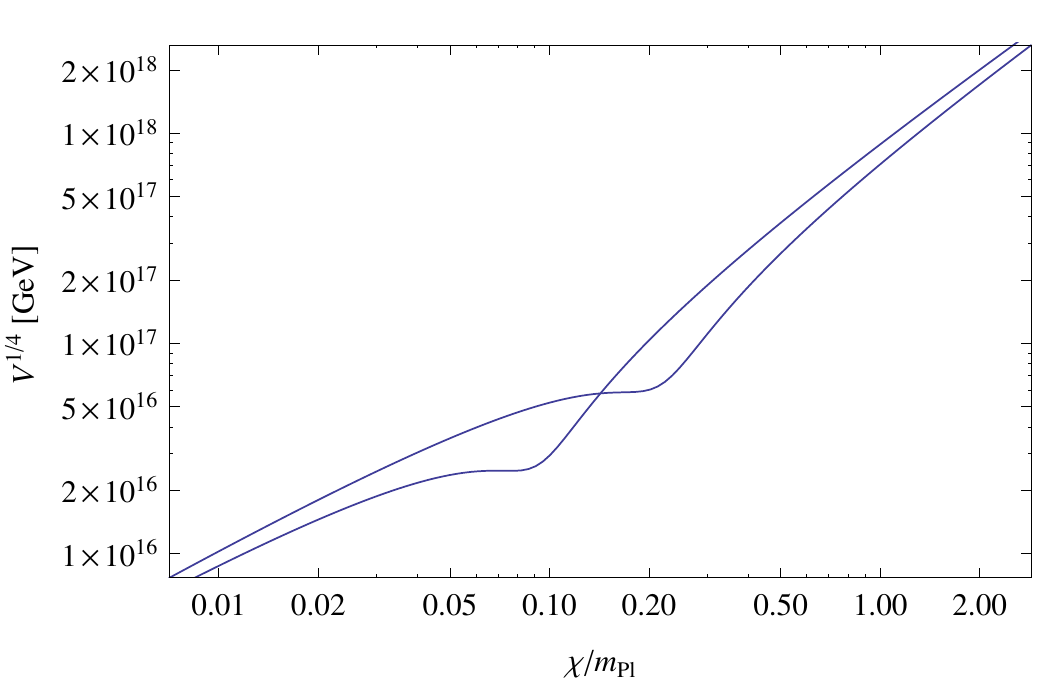}  
\caption{Higgs potential as a function of the Higgs field $\chi$ in units of $m_{Pl}=1.22\times 10^{19}$ GeV. We fixed  $\alpha_3(m_Z)=0.1184$.  The upper curve correspond to $m_H =125, m_t= 171.0305$, which shows good agreement with the results of~\cite{strumia2}, fig.~7.  The lower curve corresponds to  $m_H =123.2$, $m_t= 170.1228$, and a value of $r=0.28$.
As mentioned in the text and discussed in~\cite{Masina:2011aa}, in order to have a sizable tunneling probability through the left side, the barrier must be very low, almost as in an inflection point.}
\label{fig-Vmin}
\vskip .2 cm
\end{figure}

Using the recent state-of-the-art theoretical calculations given in~\cite{strumia2,strumia3}, we show the specific values of the top and Higgs masses
which allow for the presence of a false minimum. Such calculations are now so accurate that also the precise value of $\alpha_3(m_Z)$ inside the present allowed experimental range $\alpha_3=0.1184\pm0.007$~\cite{PDG} is a relevant parameter.
As an example, in fig.\ref{fig-Vmin} we also show the Higgs potential for very specific values of $m_t$ and $m_H$, showing agreement with fig.7 of~\cite{strumia2}.
The extremely precise values shown in the caption are not to be taken sharply, because of a theoretical 
uncertainty, estimated to be of about $0.3$  GeV on $m_H$ and about $0.15$ GeV on $m_t$.

\begin{figure*}[t!]
 \centering 
\includegraphics[width=9cm]{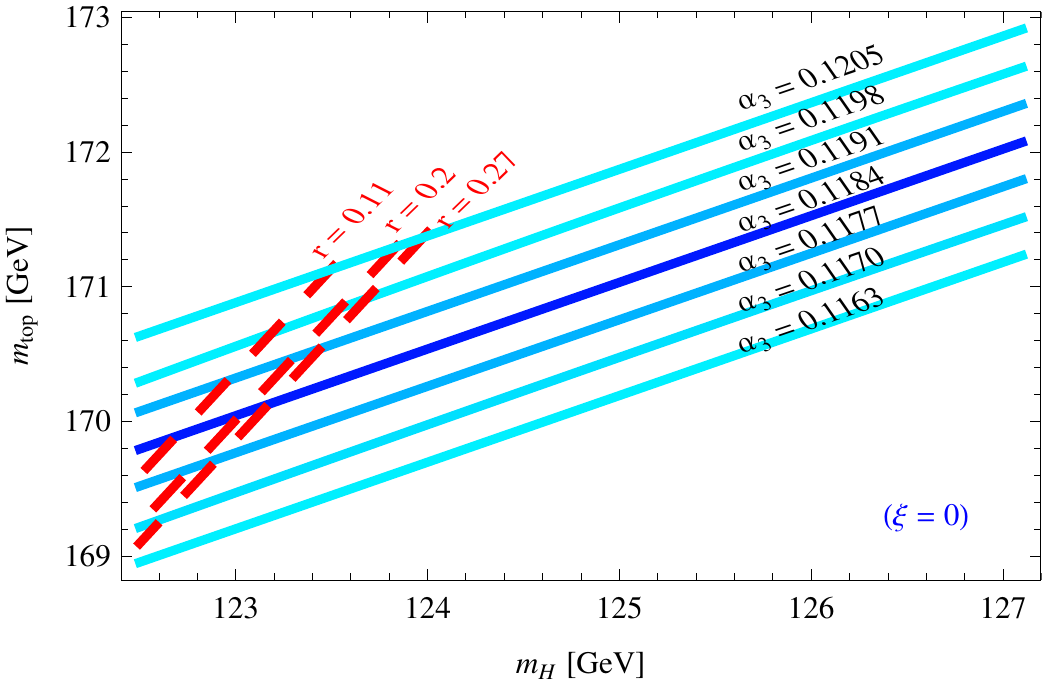}
\caption{ The solid lines indicate the $m_t-m_H$ values compatible with a shallow Higgs false minimum, taking a 3$\sigma$ variation of $\alpha_s(m_Z)$ around its central value 
of 0.1184. Here the Higgs is assumed to be minimally coupled ($\xi$=0).
There is a (vertical) uncertainty  of about $0.15$ GeV in $m_t$ and a (horizontal) one of $0.3$ GeV in $m_H$ due to the present~\cite{strumia2,strumia3} theoretical uncertainties in the RGE. The red dashed lines represent the values of the tensor-to-scalar ratio associated to the energy scale of the minimum, where we displayed the $r=0.2^{+0.07}_{0.09}$ values which correspond to the claimed recent detection by BICEP2~\cite{BICEP2}.}
\label{fig-mtmh}
\vskip .22 cm
\end{figure*}

Increasing (decreasing)  $m_t$, one has also to increase (decrease)  $m_H$ in order to develop the shallow false minimum; 
accordingly, the value of both $V(\chi_0)$ and $\chi_0$ increase (decrease).
The solid blue lines in fig.\ref{fig-mtmh} show the points in the $m_t-m_H$ plane where the shallow SM false minima exists, for different values of $
\alpha_3(m_Z)$ in the 3-sigma range.
The red dashed lines show the regions in ($m_H, m_t$) which corresponds to a value of $r$ compatible with observations, directly given by the height of  the potential via the equation
$r=\frac{V(\chi_0)}{(2.24\times 10^{16} {\rm GeV})^4} 0.2$ and it turns out that these lines point to a too small value of $m_H$ compared to the present experimental values. Of course this is a valid conclusion under the assumption that the theoretical errors on $m_H$ are now estimated to be of about $0.3$ GeV by~\cite{strumia2,strumia3}. This scenario could still be compatible with data only if the theoretical errors in the RGE and the matching conditions were larger, say of about $1$ GeV.

\begin{figure*}[t!]
\includegraphics[width=7.5cm,height=6cm]{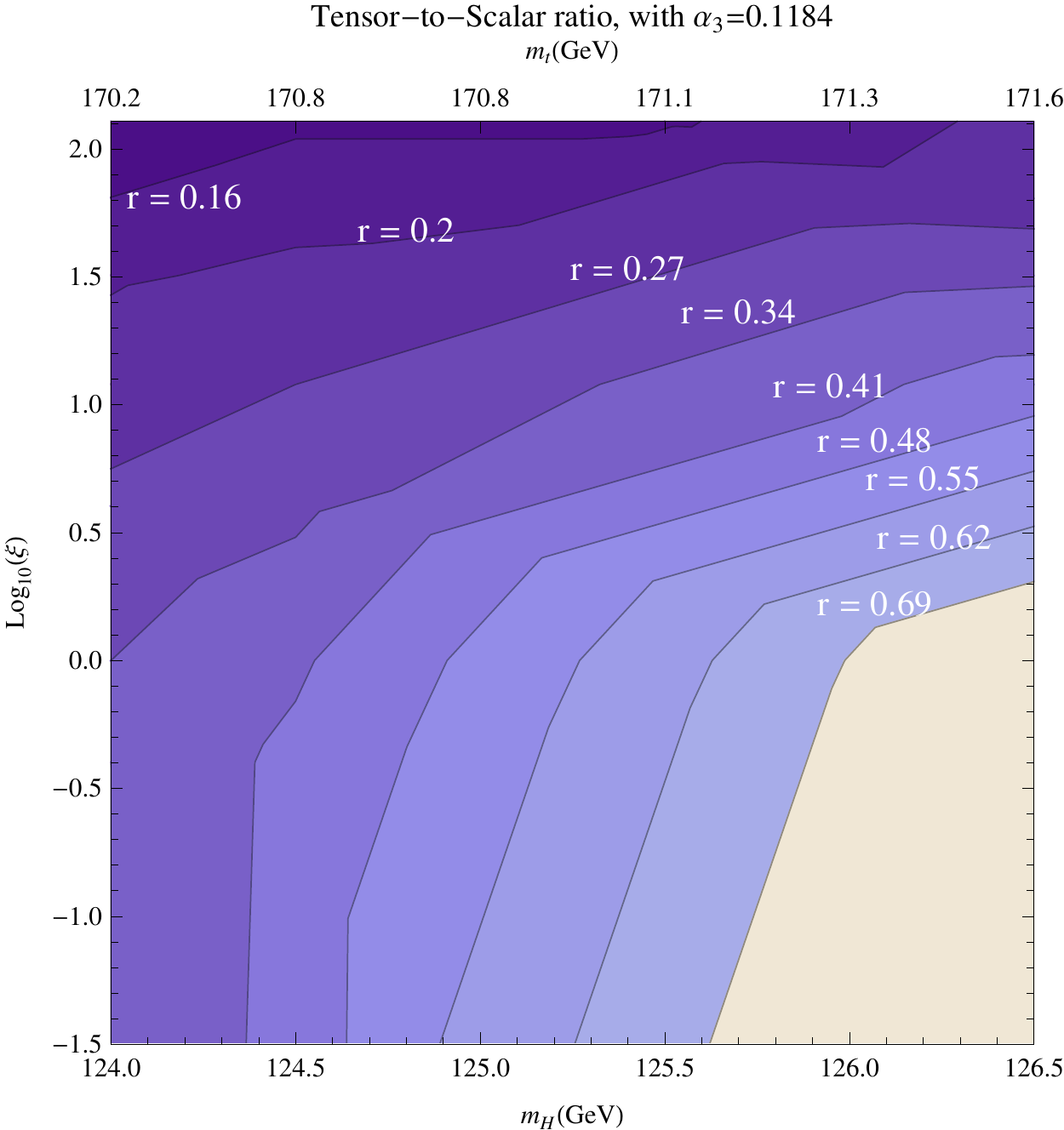}
\includegraphics[width=7.5cm,height=6cm]{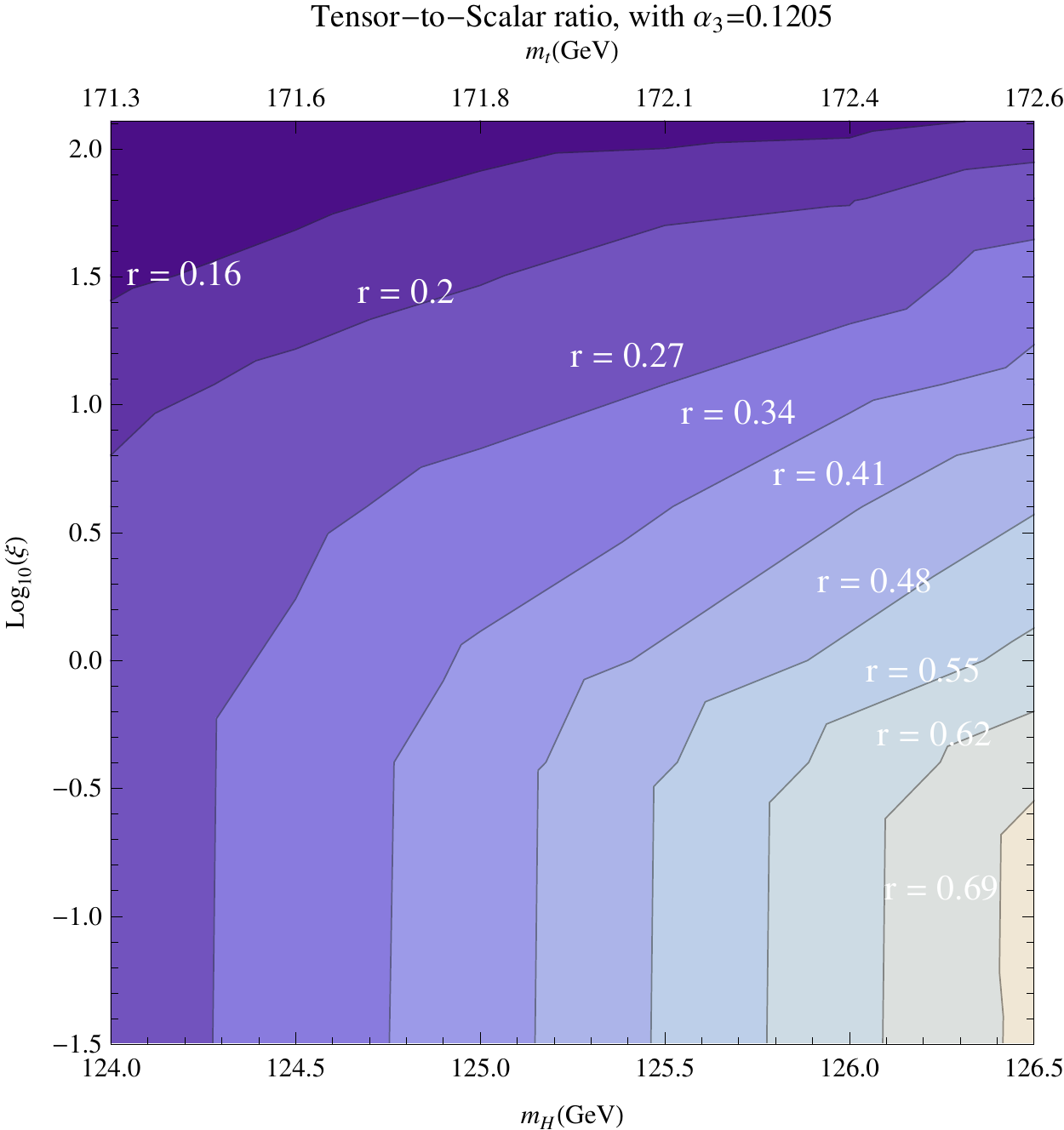}
\caption{ The contours indicate different values of the tensor-to-scalar ratio $r$ associated to the energy scale of the minimum when introducing a the nonminimal coupling $\xi$; we show it as a function of the Higgs and top mass for the central value of $\alpha_3$ (left plot) and for a larger value (3$\sigma$ deviation, right plot). There is a (vertical) uncertainty  of about $0.15$ GeV in $m_t$ and a (horizontal) one of $0.3$ GeV in $m_H$ due to the present~\cite{strumia2,strumia3} theoretical uncertainties in the RGE.}
\label{plotxi}
\vskip .22 cm
\end{figure*}

Discarding such possibility we show that this could be cured by introducing a non-minimal coupling $\xi \chi^2 R$ also for the Higgs field. This transforms the potential in the slow roll phase in the Einstein frame in the following form (see~\cite{hep-ph/0511207} and citations therein):
$$
V(\chi)= \frac{\lambda_{\rm eff}/4 \chi^4}{(1+\xi (\chi^2/M^2))^2}
$$
which suppresses the potential at large field values and we show in fig.~\ref{plotxi} that a moderately large value of $\xi$ allows for the potential to  fit CMB observations, also for larger $m_H$, compatible with the LHC measurements. However let us comment that this is based only on a tree level analysis and when having a value of the extra coupling constant $\xi$ larger than  ${\cal O}(1)$ we have to be careful about the regime of validity of the theory, similarly to the models proposed in~\cite{Bezrukov:2007ep}.

\begin{figure}[t!]
 \centering 
 \includegraphics[width=9cm]{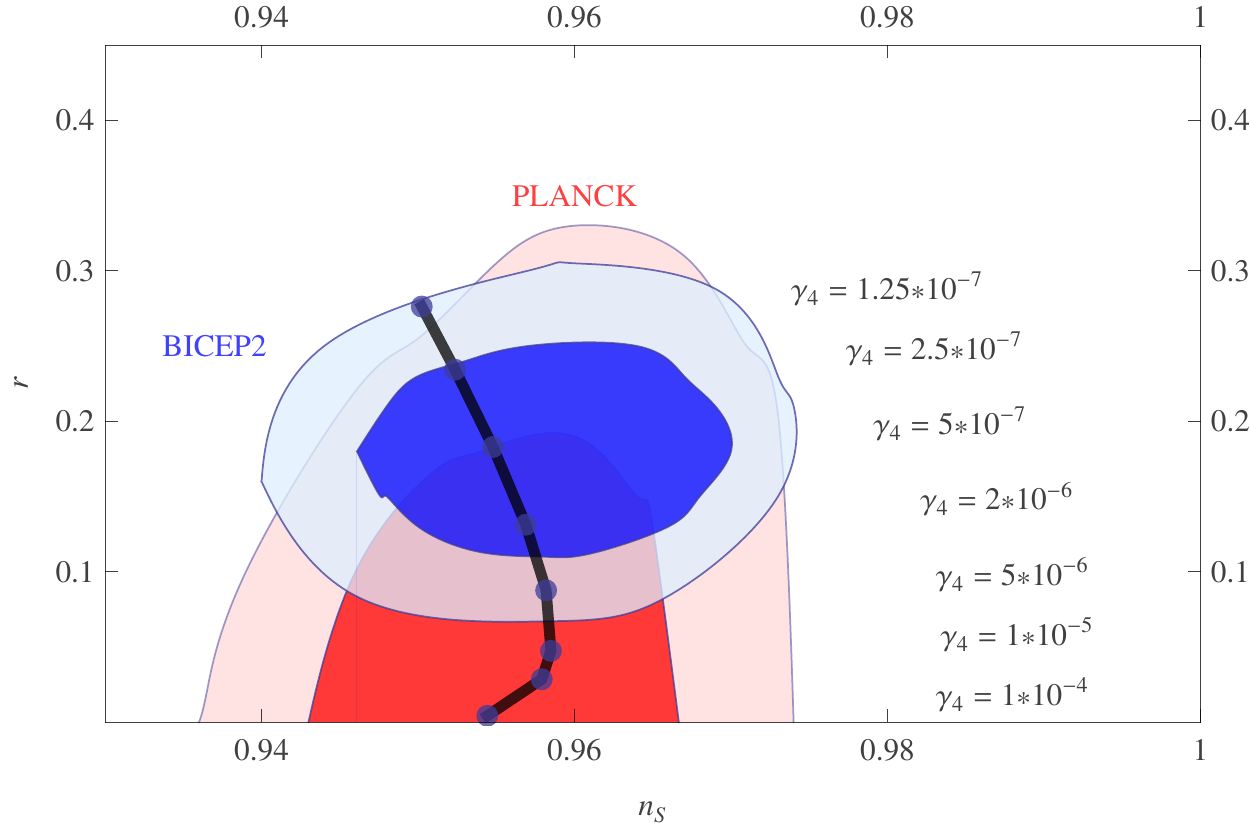}  
\caption{Spectral index $n_S$ and tensor-to-scalar ratio $r$ for the scalar-tensor model of~\cite{Masina:2011aa}. Here the number of efolds is assumed to be  $\bar{N}=60$. Only one nonvanishing parameter $\gamma_4$ was assumed to be present in eq.~\ref{eq-ftot} and the various points along the black lines correspond to different values of $\gamma_4$. }
\label{nsr}
\vskip .2 cm
\end{figure}

Finally we also show the results for the $n_s-r$ plot for the scalar tensor model proposed in~\cite{Masina:2011aa} in figure~\ref{nsr}. Such a model is based on the introduction of an additional non-minimally coupled field $\phi$ with the following action:
\begin{equation}
-S=\int d^4 x \sqrt{-g} \left[ {\cal L}_{SM} + \frac{(\partial_\mu\phi\partial^\mu\phi)}{2} -\frac{M^2}{2} f(\phi) R -U(\phi) \right]  
\label{azione}
\end{equation}
where $ {{\cal L}}_{SM}$ is the Standard Model Lagrangian and 
 the potential $U$ is not specified and assumed to be relevant only after Inflation to stabilize the field. The coupling function is given by:
\begin{equation}
f( \phi)\simeq 1 +\beta \left( \frac{ \phi}{M} \right)^2  +\sum_{n \ge 4}  \gamma_n \left( \frac{ \phi}{M} \right)^n \ \,\,.
\label{eq-ftot}
\end{equation}
with the requirement that $f(\phi)>\phi^2$, which is guaranteed for instance if the couplings $\gamma_n $  are positive. Under these conditions $H(t)$ starts almost constant and after a sufficient umber of efolds starts decreasing as  a power law, allowing for a tunneling transition when $H^4$ becomes equal to the tunneling rate per unit volume $\Gamma$. As a simple example we will study the case in which $\gamma_4>0$ and all other couplings are vanishing.
If the Higgs fields is at the minimum and it is not evolving during Inflation the only time dependent quantity is $\phi$ and so we can go the Einstein frame with a metric defined by $\bar{g}_{\mu \nu}\equiv f(\phi) g_{\mu \nu}$ and study the evolution  of a canonically normalized field $\Phi$ defined through $d\Phi=d\phi \sqrt{K(\phi)}$, where
$$K(\phi)\equiv {2f(\phi)+3 M^2 f^{'2}(\phi)\over 2f^2(\phi)}
$$
In this frame the action becomes:
\be
S_{E}= {1\over 2}\int d^4x\ \sqrt{-\bar{g}}[M^2 \bar{R}-(\bar{\partial}\Phi)^2- 2 \bar{{\cal L}}_{SM} ]
\label{e-action-2} \, .
\ee
where the bar represents quantities in the Einstein frame. The Higgs potential contained in $\bar{{\cal L}}_{SM}$ is now $ V(\chi)/f(\Phi)^{2}$, so that the potential energy at the false Higgs minimum $\chi_0$ gives rise to a potential term for $\Phi$
\be
S_{E}^{\chi_0}=\int d^4x\ \sqrt{-\bar{g}} \bar{V} \,\,\,\,, \,\,\, \bar{V}\equiv  \frac{V(\chi_0)}{f(\Phi)^{2}} \,\,. 
\label{eq-Vb0}
\ee
This acts as a hill-top potential for the $\Phi$ field and we assume that $\Phi$ rolls down the potential from small to high values. 
Given the potential we define as usual the slow-roll parameters
\be 
\epsilon(N) = \frac{1}{2} \left| \frac{1}{\bar V}  \frac{d \bar V}{d(\Phi/M) }\right|^2 
\qquad
\eta(N)= \frac{1}{\bar V}  \frac{d^2 \bar V}{d(\Phi/M)^2 }
\ee
We solve numerically the Klein-Gordon equation for $\Phi$ under such a potential  assuming that it starts as close to zero as possible ({\it i.e.} with an initial value $\Phi_I$ given by the quantum fluctuations $\Phi_I=\bar{H}/(2\pi)$) together with the Friedmann equation, which gives us a numerical solution for $\Phi(N)$, where N is the number of efolds at some value of the scale factor $\bar{a}$, defined as $N \equiv \ln(\bar{a}_F/\bar{a})$, where $\bar{a}_F$ is the scale factor at the end of the Inflationary stage, defined by either $\epsilon$ or $\eta$ becoming of order 1.
Finally we evaluate the quantites 
$$r \equiv P_T/P_S = 16 \epsilon_{\bar N} \, \qquad
n_S=1 -6 \epsilon_{\bar N} +2 \eta_{\bar N} \, $$

where ${\bar N}$ is the number of efolds that corresponds to a certain cosmological scale. We take $\bar{N}=60$ although the precise number depends on the history of the evolution subsequent to Inflation. The results are shown in fig.\ref{nsr}.\footnote{Note that in ref.\cite{Masina:2011aa} the calculations were obtained in an expansion for $\Phi\ll M$ for the slow-roll parameters, which does not hold for very small $\gamma_4$, therefore leading to incorrect results for $\gamma_4\ll 10^{-4}$, which corresponds to a sizable $r$ in fig.\ref{nsr}.}

In conclusion we have shown that a mechanism in which the energy density for Inflation is provided by  the Higgs field starting in a  shallow false minimum can be compatible with observations only if $m_H\lesssim 124$ GeV (and also $m_t$ has to be smaller than about $171$ GeV ). This is significantly lower than experimental observations unless the theoretical error on the RGE, of about $0.3$ GeV on $m_H$ according to~\cite{strumia2,strumia3}, is largely underestimated. We have also shown that introducing a non-minimal coupling of the Higgs of about ${\cal O}(10)$ can push this upper bound on larger values, closer to  present observations.
Of course such bounds can be also evaded by modifying the shape of the potential in a model-dependent way. For instance a possibility would be to introduce direct couplings of the Higgs with the extra scalar as in~\cite{strumiascalar}, which would change the potential either at tree level or in the RGE, or perhaps to allow a faster change of the barrier in hybrid models during Inflation.

%
%

%
%
%
%

\section*{Acknowledgements}
We thank A. Strumia and F.~Bezrukov for useful discussions.


\end{document}